\documentclass[numreferences]{kapproc} 
\normallatexbib

\usepackage{amssymb}
\usepackage[dvips]{graphicx}
\begin{document}

\articletitle{Polarons in complex oxides and molecular nanowires}

\author{A. S. Alexandrov}
\affil{Department of Physics, Loughborough University,\\
Loughborough, United Kingdom} \email{a.s.alexandrov@lboro.ac.uk}

\begin{abstract}
There is a growing understanding that transport properties of
complex oxides and  individual molecules are dominated by polaron
physics. In superconducting oxides the long-range Fr\"{o}hlich and
short-range Jahn-Teller electron-phonon interactions bind carriers
into real space pairs - small bipolarons with surprisingly low
mass but sufficient binding energy, while the long-range Coulomb
repulsion keeps bipolarons apart preventing their clustering. The
bipolaron theory numerically explains high Tc values without any
fitting parameters and describes other key features of the
cuprates. The same approach provides  a new insite into the theory
of transport through molecular nanowires and quantum dots (MQD).
Attractive polaron-polaron correlations lead to a "switching"
phenomenon in the current-voltage characteristics of MQD. The
degenerate MQD with strong electron-vibron coupling has two stable
current states (a volatile memory), which might be useful in
molecular electronics.
\end{abstract}


\section*{Introduction}
When  the electron-phonon (e-ph) interaction energy $E_{p}$ is
larger than  their kinetic energy, electrons in the Bloch band are
``dressed'' by phonons.  If phonon frequencies are very low, the
local lattice deformation  traps the electron even in a perfect
crystal lattice. This {\it self-trapping }phenomenon was predicted
by Landau \cite{lan2}. It has been studied in greater detail by
Pekar \cite{pek}, Fr\"{o}hlich \cite{fro2}, Feynman \cite{fey0},
Devreese \cite{dev} and other authors in the effective mass
approximation, which leads to the so-called {\it large} or {\it
continuous} polaron. The large polaron propagates through the
lattice as a free electron but with an enhanced effective mass.

In the strong-coupling regime, $\lambda=E_p/D
>1,$ the finite bandwidth $2D$ becomes important, so that the effective
mass approximation cannot be applied. The electron is called a
{\it small} or {\it lattice} polaron in this regime. The
self-trapping is never ``complete'', that is any polaron can
tunnel through the lattice. Only in the extreme {\it adiabatic}
limit, when the phonon frequencies tend to zero, the self-trapping
is complete, and the polaron motion is no longer translationally
continuous. The main features of the small polaron were understood
by Tjablikov \cite{tja}, Yamashita and Kurosava \cite{yam}, Sewell
\cite{sew}, Holstein \cite{hol} and his school \cite{frihol,emi},
Lang and Firsov \cite{fir}, Eagles \cite {eag}, and others and
described in several review papers and textbooks \cite
{dev,app,fir2,bry,mah,alemot}. An exponential reduction of the
bandwidth at large values of $\lambda $ and phonon side-bands are
among  those features.

The lattice deformation also strongly affects the interaction
between electrons. At large distances polarons repel each other in
ionic crystals, but their Coulomb repulsion is substantially
reduced
due to the ion polarization. Nevertheless two $%
large$ polarons can be bound into a $large$ bipolaron by an
exchange interaction even with no additional e-ph interaction but
the Fr\"{o}hlich one \cite{dev}.

When a short-range deformation potential and molecular-type (i.e.
Jahn-Teller \cite{mul00}) e-ph interactions are taken into account
together with the Fr\"{o}hlich interaction \cite{ale5}, they can
overcome the Coulomb repulsion. The resulting  interaction becomes
attractive at a short distance of about a lattice constant. Then
two  small polarons easily form a bound state, i.e. a $small$
bipolaron, because their band is  narrow. Consideration of
particular lattice structures shows that small bipolarons are
mobile even when the electron-phonon coupling is strong and the
bipolaron binding energy is large \cite{ale5}. Hence the polaronic
Fermi liquid transforms into a Bose liquid of double-charged
carriers in the strong-coupling regime. The Bose-liquid is stable
because bipolarons repel each other \cite{ale5}. Here we encounter
a novel electronic state of matter, a charged Bose liquid,
qualitatively different from the normal Fermi-liquid and from the
BCS superfluid.

  Experimental  evidence
for an exceptionally strong electron-phonon interaction in high
temperature superconductors is now  overwhelming. As we discussed
in detail elsewhere \cite{book},  the extension of the BCS theory
towards the strong interaction between electrons and ion
vibrations  describes the phenomenon naturally. High temperature
superconductivity exists in the crossover region of the
electron-phonon interaction strength from the BCS-like to
bipolaronic superconductivity as was predicted before \cite{ale0},
and explored in greater detail by many authors after the discovery
\cite{workshop}.

Small polarons with their phonon side-bands and attractive
correlations are quite feasible also in molecular nanowires and
quantum dots (MQD) used as the ``transmission lines''
\cite{lehn90,tour00} and active molecular elements
\cite{mark98,pat99} in molecular-scale electronics \cite{mark98}.
It has been experimentally demonstrated that the low-bias
conductance of molecules is dominated by resonant tunneling
through coupled electronic and vibration levels \cite{zhit02}.
Conductance peaks due to electron-vibron interactions has been
seen in C$_{60}$ \cite{par}. Different aspects of the
electron-phonon/vibron (e-ph) interaction effect on the tunneling
through molecules and quantum dots (QD) have been studied by
several authors \cite{win2,li,kan,erm,ven,fis,lun}. In particular,
Glazman and Shekhter, and later Wingreen {\it et al.}\cite{win2}
presented the exact resonant-tunneling transmission probability
fully taking into account the e-ph interaction on a nondegenerate
resonant site. Phonons produced transmission side-bands but did
not affect the integral transmission probability. Li, Chen and
Zhou \cite{li} studied the conductance of a double degenerate (due
to spin) quantum dot with Coulomb repulsion and the e-ph
interaction. Their numerical results also showed the side-band
peaks and the main peak related to the Coulomb repulsion, which
was decreased by the e-ph interaction. Kang \cite{kan} studied the
boson (vibron) assisted transport through a double-degenerate QD
coupled to two {\it superconducting} leads and found multiple
peaks in the I-V curves, which originated from the singular BCS
density of states and the phonon side-bands.

  While a correlated transport
through mesoscopic systems with repulsive electron-electron
interactions  received considerable interest in the past, and
continues to be the focus of intense investigations \cite{here},
 much less has been known about a role of attractive
correlations in MQD. Recently we have proposed a negative$-U$
Hubbard model of a $d$-fold degenerate quantum dot
\cite{alebrawil}. We argued that the {\em attractive} electron
correlations  caused by a strong electron-phonon (vibron)
interaction in the molecule, and/or by the valence fluctuations
provide a molecular switching effect, when the current-voltage
(I-V) characteristics show two branches with high and low current
for the same voltage. The effect was observed in a few
experimental studies with complex \cite{pat99} and  simple
molecules \cite{exp}.

Here we review the analytical theory of a {\it correlated}
transport through a degenerate molecule quantum dot (MQD) fully
taking into account both Coulomb and e-ph interactions
\cite{alebra}. We show that the phonon side-bands significantly
modify the switching behavior of the I-V curves in comparison with
the negative-$U$ Hubbard model \cite{alebrawil}. Nevertheless, the
switching effect is robust. It shows up when the effective
interaction of polarons is attractive and the state of the dot is multiply degenerate, $d>2$.%

\section{Attractive correlations of small polarons}

Employing the canonical polaron formalism with a generic
``Fr\"{o}hlich-Coulomb'' Hamiltonian, allows us explicitly
calculate the effective attraction of small polarons
\cite{alekor2}. The Hamiltonian includes the infinite-range
Coulomb, $V_c$ and electron-phonon interactions. The implicitly
present infinite on-site repulsion (Hubbard $U$) prohibits double
occupancy and removes the need to distinguish the fermionic spin.
Introducing spinless fermion operators $c_{{\bf n}}$ and phonon
operators $d_{{\bf m}\nu }$, the Hamiltonian is written as
\begin{eqnarray}
H &=&\sum_{{\bf n\neq n^{\prime }}}T({\bf n-n^{\prime }})c_{{\bf n}%
}^{\dagger }c_{{\bf n^{\prime }}}+\sum_{{\bf n\neq n^{\prime }}}V_{c}({\bf %
n-n^{\prime }})c_{{\bf n}}^{\dagger }c_{{\bf n}}c_{{\bf n^{\prime }}%
}^{\dagger }c_{{\bf n^{\prime }}}+ \\
&&\omega _{0}\sum_{{\bf n\neq m,}\nu }g_{\nu }({\bf m-n})({\bf
e}_{\nu }\cdot {\bf e}_{{\bf m-n}})c_{{\bf n}}^{\dagger }c_{{\bf
n}}(d_{{\bf m}\nu
}^{\dagger }+d_{{\bf m}\nu })+  \nonumber \\
&&\omega _{0}\sum_{{\bf m},\nu }\left( d_{{\bf m}\nu }^{\dagger }d_{{\bf m}%
\nu }+\frac{1}{2}\right) .  \nonumber
\end{eqnarray}
The e-ph term is written in real space, which is more convenient
in working with complex lattices.

In general, the many-body model Eq.(1) is of considerable
complexity. However, we are interested in the limit of the strong
e-ph interaction. In this case, the kinetic energy is a
perturbation and the model can be grossly simplified using the
canonical transformation \cite{fir} in the Wannier representation
for electrons and phonons,
\[
S=\sum_{{\bf m\neq n,}\nu }g_{\nu }({\bf m-n})({\bf e}_{\nu }\cdot {\bf e}_{%
{\bf m-n}})c_{{\bf n}}^{\dagger }c_{{\bf n}}(d_{{\bf m}\nu }^{\dagger }-d_{%
{\bf m}\nu }).
\]

The transformed Hamiltonian is
\begin{eqnarray}
\tilde{H} &=&e^{-S}He^{S}=\sum_{{\bf n\neq n^{\prime }}}\hat{\sigma}_{{\bf %
nn^{\prime }}}c_{{\bf n}}^{\dagger }c_{{\bf n^{\prime }}}+\omega _{0}\sum_{%
{\bf m}\alpha }\left( d_{{\bf m}\nu }^{\dagger }d_{{\bf m}\nu }+\frac{1}{2}%
\right) + \\
&&\sum_{{\bf n\neq n^{\prime }}}v({\bf n-n^{\prime }})c_{{\bf
n}}^{\dagger
}c_{{\bf n}}c_{{\bf n^{\prime }}}^{\dagger }c_{{\bf n^{\prime }}}-E_{p}\sum_{%
{\bf n}}c_{{\bf n}}^{\dagger }c_{{\bf n}}.  \nonumber
\end{eqnarray}
The last term describes the energy gained by polarons due to e-ph
interaction. $E_{p}$ is the familiar polaron level shift
\begin{equation}
E_{p}=\omega_0 \sum_{{\bf m}\nu }g_{\nu }^{2}({\bf m-n})({\bf
e}_{\nu }\cdot {\bf e}_{{\bf m-n}})^{2},
\end{equation}
which is independent of ${\bf n}$. The third term on the
right-hand side in Eq.(2) is the polaron-polaron interaction:
\begin{equation}
v({\bf n-n^{\prime }})=V_{c}({\bf n-n^{\prime }})-V_{ph}({\bf n-n^{\prime }}%
),
\end{equation}
where
\begin{eqnarray*}
V_{ph}({\bf n-n^{\prime }}) &=&2\omega _{0}\sum_{{\bf m,}\nu }g_{\nu }({\bf %
m-n})g_{\nu }({\bf m-n^{\prime }})\times \\
&&({\bf e}_{\nu }\cdot {\bf e}_{{\bf m-n}})({\bf e}_{\nu }\cdot {\bf e}_{%
{\bf m-n^{\prime }}}).
\end{eqnarray*}
The phonon-induced interaction $V_{ph}$ is due to displacements of
common
ions by two electrons. Finally, the transformed hopping operator $\hat{\sigma%
}_{{\bf nn^{\prime }}}$ in the first term in Eq.(2) is given by
\begin{eqnarray}
\hat{\sigma}_{{\bf nn^{\prime }}} &=&T({\bf n-n^{\prime }})\exp \left[ \sum_{%
{\bf m,}\nu }\left[ g_{\nu }({\bf m-n})({\bf e}_{\nu }\cdot {\bf e}_{{\bf m-n%
}})\right. \right. \\
&&-\left. \left. g_{\nu }({\bf m-n^{\prime }})({\bf e}_{\nu }\cdot {\bf e}_{%
{\bf m-n^{\prime }}})\right] (d_{{\bf m}\alpha }^{\dagger
}-d_{{\bf m}\alpha })\right] .  \nonumber
\end{eqnarray}
This term is a perturbation at large $\lambda $. It is absent in
 an isolated MQD, modeled as a single degenerate atomic
level, so that the canonical transformation solves the problem
exactly for any number of electrons (see below).  In a crystal the
term allows for a bipolaron tunnelling and high temperature
superconductivity \cite{ale0}. In  particular crystal structures
like perovskites, the tunnneling appears already in the first
order in $T({\bf n})$, so that $\hat{\sigma}_{{\bf nn^{\prime }}}$
can be averaged over phonons. The result is
\begin{equation}
t({\bf n-n^{\prime }})\equiv \left\langle \left\langle \hat{\sigma}_{{\bf %
nn^{\prime }}}\right\rangle \right\rangle _{ph}=T({\bf n-n^{\prime
}})\exp [-g^{2}({\bf n-n^{\prime }})],
\end{equation}
\begin{eqnarray*}
g^{2}({\bf n-n^{\prime }}) &=&\sum_{{\bf m},\nu }g_{\nu }({\bf m-n})({\bf e}%
_{\nu }\cdot {\bf e}_{{\bf m-n}})\times \\
&&\left[ g_{\nu }({\bf m-n})({\bf e}_{\nu }\cdot {\bf e}_{{\bf
m-n}})-g_{\nu
}({\bf m-n^{\prime }})({\bf e}_{\nu }\cdot {\bf e}_{{\bf m-n^{\prime }}})%
\right] .
\end{eqnarray*}
By comparing Eqs.(6) and Eqs.(3,4), the mass renormalization
exponent can be expressed via $E_{p}$ and $V_{ph}$ as follows
\begin{equation}
g^{2}({\bf n-n^{\prime }})=\frac{1}{\omega _{0}}\left[ E_{p}-\frac{1}{2}%
V_{ph}({\bf n-n^{\prime }})\right] .
\end{equation}

When $V_{ph}$ is larger than $V_{c}$ the full interaction becomes
negative and polarons form pairs. The real space representation
allows us to elaborate more physics behind the lattice sums in
Eq.(3) and Eq.(4). If a carrier (electron or hole) acts on an ion
with a force ${\bf f}$, it displaces the ion by some vector ${\bf
x}={\bf f}/s$. Here $s$ is the ion's
force constant. The total energy of the carrier-ion pair is $-{\bf f}%
^{2}/(2s)$. This is precisely the summand in Eq.(3) expressed via
dimensionless coupling constants. Now consider two carriers
interacting with
the {\em same} ion, see Fig.1a. The ion displacement is ${\bf x}=({\bf f}%
_{1}+{\bf f}_{2})/s$ and the energy is $-{\bf f}_{1}^{2}/(2s)-{\bf f}%
_{2}^{2}/(2s)-({\bf f}_{1}\cdot {\bf f}_{2})/s$. Here the last
term should be interpreted as an ion-mediated interaction between
the two carriers. It depends on the scalar product of ${\bf
f}_{1}$ and ${\bf f}_{2}$ and consequently on the relative
positions of the carriers with respect to the ion. If the ion is
an isotropic harmonic oscillator, as we assume here, then the
following simple rule applies. If the angle $\phi $ between ${\bf
f}_{1}$ and ${\bf f}_{2}$ is less than $\pi /2$ the
polaron-polaron interaction will be attractive, if otherwise it
will be repulsive. In general, some ions will generate attraction,
and some repulsion between polarons, Fig. 1b.

\begin{figure}[tbp]
\begin{center}
\includegraphics[angle=-0,width=0.57\textwidth]{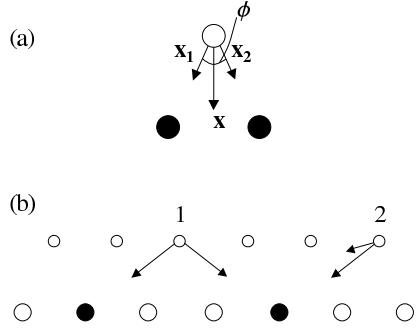} \vskip -0.5mm
\end{center}
\caption{The mechanism of polaron-polaron interaction. (a)
Together, the two polarons (solid circles) deform the lattice more
effectively than separately. An effective attraction occurs when
the angle $\phi $ is less than $\pi /2$ . (b) A mixed situation.
Ion 1 results in repulsion between two polarons while ion 2
results in attraction. }
\end{figure}
The overall sign and magnitude of the interaction is given by the
lattice sum in Eq.(4), the evaluation of which is elementary. One
should also note that according to Eq.(7) an attractive
interaction reduces the polaron mass (and consequently the
bipolaron mass), while repulsive interaction enhances the mass.

\section{Steady current through MQD}

Let us now  apply the polaron formalism to MQD \cite{alebra}. Here
bipolarons might not exist because of a finite lifetime of
electrons on a molecule connected with the leads, but the
attractive correlations could strongly modify the current-voltage
characteristics. We employ the Landauer-type expression for a
steady current through a region of interacting electrons, derived
by Meir and Wingreen \cite{meir} as (in units $\hbar =k_{B}=1$)
\begin{equation}
I(V)=-{\frac{e}{\pi }}\int_{-\infty }^{\infty }d\omega \left[
f_{1}(\omega
)-f_{2}(\omega )\right] {\rm ImTr}\left[ \hat{\Gamma}(\omega )\hat{G}%
^{R}(\omega )\right] ,  \label{eq:Igen}
\end{equation}
where $f_{1(2)}(\omega )=\left\{ \exp [(\omega +\Delta \mp
eV/2)/T]+1\right\} ^{-1},$  $T$ is the temperature, $\Delta $ is
the position of the lowest unoccupied molecular level with respect
to the chemical potential. $\hat{\Gamma}(\omega )$ depends on the
density of states (DOS) in the leads and on the hopping integrals
connecting one-particle states in the left\ (1) and the right (2)
leads with the states in MQD, Fig.~2.

\begin{figure}[tbp]
\begin{center}
\includegraphics[angle=-0,width=0.37\textwidth]{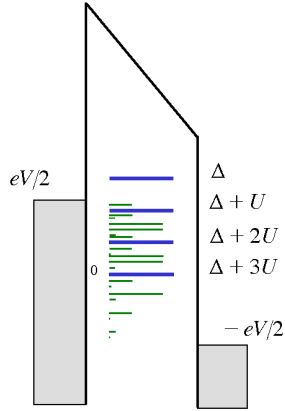} \vskip -0.5mm
\end{center}
\caption{Schematic of the energy levels and phonon side-bands for
molecular quantum dot under bias voltage $V$ ($eV/2\Delta=0.75$)
with the coupling constant $\gamma^2=11/13$. The level is assumed
to be 4-fold degenerate $(d=4)$ with energies $\Delta+ rU$, $r=0,
\dots, (d-1)$ (thick bars). Thin bars show the vibron side-bands
with the size of the bar proportional to the weight of the
particular contribution in the density of states (see text) in the
case of one vibron with frequency $\omega_0/\Delta=0.2$ at $T=0$.
Only the bands in the energy window $(eV/2, -eV/2)$ (shown) will
contribute to current at zero temperature.}
\end{figure}

This formula includes, by means of the Fourier transform of the
full molecular retarded Green's function (GF), $\hat{G}^{R}(\omega
),$ the e-ph and Coulomb interactions inside the MQD and coupling
to the leads. Since the leads are metallic, electron-electron and
e-ph interactions in the leads, and interactions of electrons in
the leads with electrons and phonons in the MQD can be neglected.
We are interested in the tunneling near the conventional
threshold, $eV=2\Delta $, Fig.2, within a voltage range about an
effective attractive potential $|U|$ caused by phonons/vibrons
(see below).

The attractive energy is the difference of two large interactions,
the
Coulomb repulsion and the phonon mediated attraction, of the order of $1{\rm %
eV}$ each. Hence, $|U|$ is of the order of a few tens of one eV.
We neglect the energy dependence of $\ \hat{\Gamma}(\omega
)\approx \Gamma $ on this scale, and assume that the coupling to
the leads is weak, $\Gamma \ll $ $|U|. $ In this case
$\hat{G}^{R}(\omega )$ does not depend on the leads. Moreover
we assume that there is a complete set of one-particle molecular states $%
\left| \mu \right\rangle $, where $\hat{G}^{R}(\omega )$ is
diagonal. With these assumptions we can reduce Eq.(\ref{eq:Igen})
to

\begin{equation}
I(V)=I_{0}\int_{-\infty }^{\infty }d\omega \left[ f_{1}(\omega
)-f_{2}(\omega )\right] \rho (\omega ),  \label{eq:Is}
\end{equation}
allowing for a transparent analysis of essential physics of the
switching phenomenon. Here $I_{0}=e\Gamma $ and the molecular DOS,
$\rho (\omega ),$ is given by
\begin{equation}
\rho (\omega )=-\frac{1}{\pi }\sum_{\mu }%
\mathop{\rm Im}%
\hat{G}_{\mu }^{R}(\omega ),
\end{equation}
where $\hat{G}_{\mu }^{R}(\omega )$ is the Fourier transform of $\hat{G}%
_{\mu }^{R}(t)=-i\theta (t)\left\langle \left\{ c_{\mu }(t),c_{\mu
}^{\dagger }\right\} \right\rangle ,$ $\left\{ \cdots ,\cdots
\right\} $ is the anticommutator, $c_{\mu }(t)=e^{iHt}c_{\mu
}e^{-iHt},$ $\theta (t)=1$ for $t>0$ and zero otherwise.

We calculate $\rho (\omega )$ exactly( see below) in the framework
of the Hamiltonian, which includes both the Coulomb $U^{C}$ and
e-ph interactions as
\begin{eqnarray}
H &=&\sum_{_{\mu }}\varepsilon _{_{\mu }}\hat{n}_{_{\mu }}+\frac{1}{2}%
\sum_{_{\mu }\neq \mu ^{\prime }}U_{\mu \mu ^{\prime }}^{C}\hat{n}_{_{\mu }}%
\hat{n}_{\mu ^{\prime }}  \nonumber \\
&&+\sum_{\mu ,q}\hat{n}_{_{\mu }}\omega _{q}(\gamma _{\mu
q}d_{q}+H.c.)+\sum_{q}\omega _{q}(d_{q}^{\dagger }d_{q}+1/2).
\end{eqnarray}
Here $\varepsilon _{_{\mu }}$ are one-particle molecular energy levels, $%
\hat{n}_{\mu }=c_{\mu }^{\dagger }c_{\mu }$ the occupation number
operators,
$c_{_{\mu }}$ and $d_{q}$ annihilates electrons and phonons, respectively, $%
\omega _{q}$ are the phonon (vibron) frequencies, and $\gamma
_{\mu q}$ are e-ph coupling constants ($q$ enumerates the vibron
modes). This Hamiltonian conserves the occupation numbers of
molecular states $\hat{n}_{_{\mu }}$. Hence it is compatible with
Eq.(\ref{eq:Is}).

\section{MQD density of states}

We apply the canonical polaron unitary transformation $e^{S}$, as
in Section 1,  integrating phonons out. The electron and phonon
operators are transformed as
\begin{equation}
\tilde{c}_{\mu }=c_{\mu }X_{\mu },
\end{equation}
and
\begin{equation}
\tilde{d}_{q}=d_{q}-\sum_{\mu }\hat{n}_{\mu }\gamma _{\mu q}^{\ast
},
\end{equation}
respectively. Here
\[
X_{\mu }=\exp \left[ \sum_{q}\gamma _{\mu q}d_{q}-H.c.\right] .
\]
The Lang-Firsov canonical transformation shifts ions to new
equilibrium positions with no effect on the phonon frequencies.
The diagonalization is exact in MQD:
\begin{equation}
\tilde{H}=\sum_{i}\tilde{\varepsilon}_{_{\mu }}\hat{n}_{\mu
}+\sum_{q}\omega _{q}(d_{q}^{\dagger
}d_{q}+1/2)+{\frac{1}{{2}}}\sum_{\mu \neq \mu ^{\prime }}U_{\mu
\mu ^{\prime }}\hat{n}_{\mu }\hat{n}_{\mu ^{\prime }},
\end{equation}
where
\begin{equation}
U_{\mu \mu ^{\prime }}\equiv U_{\mu \mu ^{\prime
}}^{C}-2\sum_{q}\gamma _{\mu q}^{\ast }\gamma _{\mu ^{\prime
}q}\omega _{q}  \label{eq:Umm1}
\end{equation}
is the interaction of polarons comprising their interaction via
molecular deformations (vibrons) and non-vibron (e.g. Coulomb
repulsion) $U_{\mu \mu ^{\prime }}^{C}$. To simplify the
discussion, we shall assume, that the Coulomb integrals do not
depend on the orbital index, i.e. $U_{\mu \mu ^{\prime }}=U.$

The molecular energy levels are shifted by the polaron level-shift
due to a deformation well created by polaron,
\begin{equation}
\tilde{\varepsilon}_{_{\mu }}=\varepsilon _{_{\mu
}}{-}\sum_{q}|\gamma _{\mu q}|^{2}\omega _{q}.
\end{equation}
Applying the same transformation in the retarded GF we obtain
\begin{eqnarray}
G_{\mu }^{R}(t) &=&-i\theta (t)\left\langle \left\{ c_{\mu
}(t)X_{\mu
}(t),~c_{\mu }^{\dagger }X_{\mu }^{\dagger }\right\} \right\rangle  \\
&=&-i\theta (t)[\left\langle c_{\mu }(t)c_{\mu }^{\dagger
}\right\rangle
\left\langle X_{\mu }(t)X_{\mu }^{\dagger }\right\rangle   \nonumber \\
&&+\left\langle c_{\mu }^{\dagger }c_{\mu }(t)\right\rangle
\left\langle X_{\mu }^{\dagger }X_{\mu }(t)\right\rangle ],
\nonumber
\end{eqnarray}
where now electron and phonon operators are averaged over the
quantum state of the transformed Hamiltonian $\tilde{H}.$ There is
no coupling between polarons and vibrons in the transformed
Hamiltonian, so that
\begin{eqnarray}
&&\left\langle X_{\mu }(t)X_{\mu }^{\dagger }\right\rangle   \nonumber \\
&=&\exp \left[ \sum_{q}\frac{|\gamma _{\mu q}|^{2}}{\sinh
\frac{\beta \omega _{q}}{2}}\left[ \cos \left( \omega
t+i\frac{\beta \omega _{q}}{2}\right) -\cosh \frac{\beta \omega
_{q}}{2}\right] \right] ,  \label{eq:XXT}
\end{eqnarray}
where $\beta =1/T$, and $\left\langle X_{\mu }^{\dagger }X_{\mu
}(t)\right\rangle =\left\langle X_{\mu }(t)X_{\mu }^{\dagger
}\right\rangle ^{\ast }$.

Next, we introduce the $N$-particle GFs, which will necessarily
appear in the equations of motion for $\left\langle c_{\mu
}(t)c_{\mu }^{\dagger }\right\rangle $, as
\begin{equation}
G_{\mu }^{(N,+)}(t)\equiv -i\theta (t)\sum_{\mu _{1}\neq \mu
_{2}\neq ...\mu }\left\langle c_{\mu }(t)c_{\mu }^{\dagger
}\prod_{i=1}^{N-1}\hat{n}_{\mu _{i}}\right\rangle ,
\end{equation}
and
\begin{equation}
G_{\mu }^{(N,-)}(t)\equiv -i\theta (t)\sum_{\mu _{1}\neq \mu
_{2}\neq ...\mu }\left\langle c_{\mu }^{\dagger }c_{\mu
}(t)\prod_{i=1}^{N-1}\hat{n}_{\mu _{i}}\right\rangle .
\end{equation}
Then, using the equation of motion for the Heisenberg polaron
operators, we derive the following equations for the $N$-particle
GFs,
\begin{eqnarray}
i\frac{dG_{\mu }^{(N,+)}(t)}{dt} &=&\delta (t)(1-n_{\mu
})\sum_{\mu _{1}\neq
\mu _{2}\neq ...\mu }\prod_{i=1}^{N-1}n_{\mu _{i}}  \nonumber \\
&&+[\tilde{\varepsilon}_{_{\mu }}+(N-1)U]G_{\mu
}^{(N,+)}(t)+UG_{\mu }^{(N+1,+)}(t),
\end{eqnarray}
and
\begin{eqnarray}
i\frac{dG_{\mu }^{(N,-)}(t)}{dt} &=&\delta (t)n_{\mu }\sum_{\mu
_{1}\neq \mu
_{2}\neq ...\mu }\prod_{i=1}^{N-1}n_{\mu _{i}}  \nonumber \\
&&+[\tilde{\varepsilon}_{_{\mu }}+(N-1)U]G_{\mu
}^{(N,-)}(t)+UG_{\mu }^{(N+1,-)}(t),
\end{eqnarray}
where $n_{\mu }=\left\langle c_{\mu }^{\dagger }c_{\mu
}\right\rangle $ is the expectation number of electrons on the
molecular level $\mu $.

We  readily solve this set of coupled equations for MQD with one
$d$-fold degenerate energy level and with the e-ph coupling
$\gamma _{\mu q}=\gamma _{q},$ which does not break the
degeneracy. Assuming that $n_{\mu }=n,$ Fourier transformation of
the set yields for $N=1$
\begin{equation}
G_{\mu }^{(1,+)}(\omega
)=(1-n)\sum_{r=0}^{d-1}\frac{Z_{r}(n)}{\omega -rU+i\delta },
\label{eq:G1+}
\end{equation}
\begin{equation}
G_{\mu }^{(1,-)}(\omega )=n\sum_{r=0}^{d-1}\frac{Z_{r}(n)}{\omega
-rU+i\delta }  \label{eq:G1-}
\end{equation}
where $\delta =+0$, and
\begin{equation}
Z_{r}(n)=\frac{(d-1)!}{r!(d-1-r)!}n^{r}(1-n)^{d-1-r}.
\end{equation}

In approximation, where we retain a coupling to a single mode with
the characteristic frequency $\omega _{0}$ and $\gamma _{q}\equiv
\gamma $, the {\em molecular DOS} is readily found as an imaginary
part of the Fourier transform of Eq.(17) using Eqs.(23,24) and
Eq.(18):
\begin{eqnarray}
&&\rho (\omega )={\cal
Z}d\sum_{r=0}^{d-1}Z_{r}(n)\sum_{l=0}^{\infty
}I_{l}\left( \xi \right)   \nonumber \\
&&\times \biggl[e^{\frac{\beta \omega _{0}l}{2}}\left[ (1-n)\delta
(\omega
-rU-l\omega _{0})+n\delta (\omega -rU+l\omega _{0})\right]   \nonumber \\
&&+(1-\delta _{l0})e^{-\frac{\beta \omega _{0}l}{2}}[n\delta
(\omega
-rU-l\omega _{0})  \nonumber \\
&&+(1-n)\delta (\omega -rU+l\omega _{0})]\biggr],  \label{eq:rho}
\end{eqnarray}
where
\begin{equation}
{\cal Z}=\exp \left[ -\sum_{{\bf q}}|\gamma _{q}|^{2}\coth
\frac{\beta \omega _{q}}{2}\right],
\end{equation}
 $\xi =|\gamma |^{2}/\sinh \frac{\beta \omega _{0}}{2},$
$I_{l}\left( \xi \right) $ is the modified Bessel function, and
$\delta _{lk}$ is the Kroneker symbol. The important feature of
the DOS, Eq.(\ref{eq:rho}), is its nonlinear dependence on the
occupation number $n,$ which leads to the switching effect and
hysteresis in the I-V characteristics for $d>2$, as will be shown
below. It contains full information about all possible correlation
and inelastic effects in transport, in particular, all the
vibron-assisted tunneling processes and phonon sidebands, and
describes the renormalization of hopping to the leads.

\section{Nonlinear rate equation and switching}

Generally, the electron density $n_{\mu }$ obeys an infinite set
of rate equations for many-particle GFs which can be derived in
the framework of a tunneling Hamiltonian including correlations
\cite{alebrawil}. In the case of MQD only weakly coupled with
leads one can apply the Fermi-Dirac golden rule to obtain an
equation for $n.$ Equating incoming and outgoing numbers of
electrons in MQD per unit time we obtain the self-consistent
equation for the level occupation $n$ as
\begin{eqnarray}
&&(1-n)\int_{-\infty }^{\infty }d\omega \left\{ \Gamma
_{1}f_{1}(\omega
)+\Gamma _{2}f_{2}(\omega )\right\} \rho (\omega )  \nonumber \\
&&-n\int_{-\infty }^{\infty }d\omega \left\{ \Gamma
_{1}[1-f_{1}(\omega )]+\Gamma _{2}[1-f_{2}(\omega )]\right\} \rho
(\omega ) =0  \label{eq:neq}
\end{eqnarray}
where $\Gamma _{1(2)}$ are the transition rates from left (right)
leads to MQD. Taking into account that $\int_{-\infty }^{\infty
}\rho (\omega )=d$, Eq.(\ref{eq:neq}) for the symmetric leads,
$\Gamma _{1}=\Gamma _{2},$ reduces to
\begin{equation}
2nd=\int d\omega \rho \left( \omega \right) \left(
f_{1}+f_{2}\right) ,
\end{equation}
which automatically satisfies $0\leq n \leq 1$. Explicitly, the
self-consistent equation for the occupation number is
\begin{equation}
n=\frac{1}{2}\sum_{r=0}^{d-1}Z_{r}(n)[na_{r}+(1-n)b_{r}],
\label{eq:neq1}
\end{equation}
where
\begin{eqnarray}
a_{r} &=&{\cal Z}\sum_{l=0}^{\infty }I_{l}\left( \xi \right) \biggr(e^{\frac{%
\beta \omega _{0}l}{2}}[f_{1}(rU-l\omega _{0})+f_{2}(rU-l\omega
_{0})]
\nonumber \\
&&+(1-\delta _{l0})e^{-\frac{\beta \omega
_{0}l}{2}}[f_{1}(rU+l\omega
_{0})+f_{2}(rU+l\omega _{0})]\biggr),  \label{eq:a} \\
b_{r} &=&{\cal Z}\sum_{l=0}^{\infty }I_{l}\left( \xi \right) \biggr(e^{\frac{%
\beta \omega _{0}l}{2}}[f_{1}(rU+l\omega _{0})+f_{2}(rU+l\omega
_{0})]
\nonumber \\
&&+(1-\delta _{l0})e^{-\frac{\beta \omega
_{0}l}{2}}[f_{1}(rU-l\omega _{0})+f_{2}(rU-l\omega _{0})]\biggr).
\label{eq:b}
\end{eqnarray}
The current is expressed as
\begin{equation}
j\equiv
\frac{I(V)}{dI_{0}}=\sum_{r=0}^{d-1}Z_{r}(n)[na_{r}^{\prime
}+(1-n)b_{r}^{\prime }],
\end{equation}
where
\begin{eqnarray}
a_{r}^{\prime } &=&{\cal Z}\sum_{l=0}^{\infty }I_{l}\left( \xi \right) %
\biggr(e^{\frac{\beta \omega _{0}l}{2}}[f_{1}(rU-l\omega
_{0})-f_{2}(rU-l\omega _{0})]  \nonumber \\
&&+(1-\delta _{l0})e^{-\frac{\beta \omega
_{0}l}{2}}[f_{1}(rU+l\omega
_{0})-f_{2}(rU+l\omega _{0})]\biggr),  \label{eq:at} \\
b_{r}^{\prime } &=&{\cal Z}\sum_{l=0}^{\infty }I_{l}\left( \xi \right) %
\biggr(e^{\frac{\beta \omega _{0}l}{2}}[f_{1}(rU+l\omega
_{0})-f_{2}(rU+l\omega _{0})]  \nonumber \\
&&+(1-\delta _{l0})e^{-\frac{\beta \omega
_{0}l}{2}}[f_{1}(rU-l\omega _{0})-f_{2}(rU-l\omega _{0})]\biggr).
\label{eq:bt}
\end{eqnarray}

There is only one physical ($0<n<0.5$) solution   of the rate
equation (\ref{eq:neq1}) and no switching for a nondegenerate,
$d=1$, and double-degenerate, $d=2$,  MQDs. However, the switching
appears for $d>2$. For example, for $d=4$
 the rate equation is of the fourth power in $n,$%
\begin{eqnarray}
2n &=&(1-n)^{3}[na_{0}+(1-n)b_{0}]  \nonumber \\
&&+3n(1-n)^{2}[na_{1}+(1-n)b_{1}]  \nonumber \\
&&+3n^{2}(1-n)[na_{2}+(1-n)b_{2}]  \nonumber \\
&&+n^{3}[na_{3}+(1-n)b_{3}].  \label{eq:nab}
\end{eqnarray}

\begin{figure}[tbp]
\begin{center}
\includegraphics[angle=-0,width=0.67\textwidth]{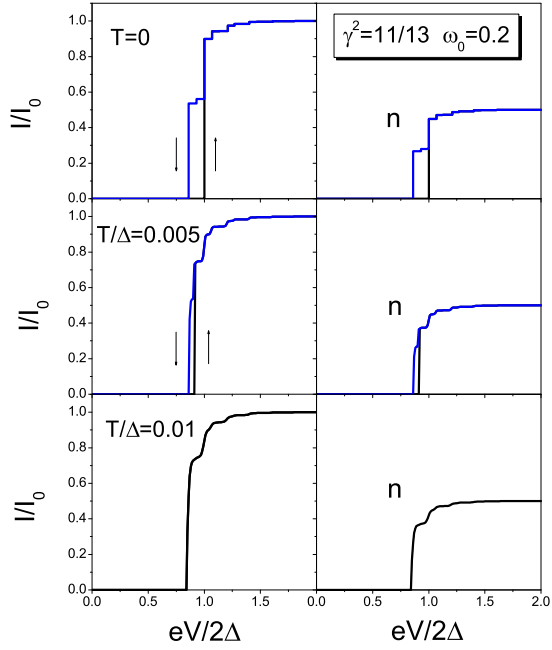} \vskip -0.5mm
\end{center}
\caption{The bistable I-V curves for tunneling through molecular
quantum dot (Fig. 2) with the electron-vibron coupling constant
$\gamma^2=11/13$ and $\omega_0/\Delta = 0.2$. The up arrows show
that the current picks up at some voltage when it is biased, and
then drops at lower voltage when the bias is being reduced. The
bias dependence of current basically repeats the shape of the
level occupation $n$ (right column). Steps on the curve correspond
to the changing population of the phonon side-bands, which are
shown in Fig. 2.}
\end{figure}

 Differently from the non-degenerate or double-degenerate MQD, the rate equation (\ref{eq:nab}) for $d=4$
  has two stable
 physical roots in a certain voltage range and the current-voltage
characteristics show a hysteretic behavior.
 We show
the numerical results for $\omega _{0}=0.2$ (in units of $\Delta
,$ as all the energies in the problem) and $U^{C}=0$ for the
coupling constant, $\gamma ^{2}=11/13$ in Fig. 3. This case
formally corresponds to a negative Hubbard $U=-2\gamma ^{2}\omega
_{0}\approx -0.4$ (we selected those values of $\gamma^2$ to avoid
accidental commensurability of  the correlated levels separated by
$U$ and the phonon side-bands). The threshold for the onset of
bistability appears at a voltage bias  $eV/2\Delta =0.86 $ for
$\gamma ^{2}=11/13$ and $\omega _{0}=0.2$).  The inelastic
tunneling processes through the level, accompanied by
emission/absorption of the vibrons, manifest themselves as steps
on the I-V curve, Figs. 3. Those steps are generated by the phonon
side-bands originating from correlated levels in the
dot with the energies $\Delta ,$ $\Delta +U,$ ..., $\Delta +(d-1)U.$ Since $%
\omega _{0}$ is not generally commensurate with $U,$ we obtain
quite irregular picture of the steps in I-V\ curves. The
bistability region shrinks down with temperature.

In conclusion, we have reviewed the multi-polaron theory of
tunneling through a molecular quantum dot (MQD) taking phonon
side-bands and attractive polaron correlations into account. The
degenerate MQD with strong electron-vibron coupling shows a
hysteretic volatile memory if the degeneracy of the molecular
level is larger than two, $d>2.$ The hysteretic behavior strongly
depends on the electron-vibron coupling and characteristic vibron
frequencies. The current bistability vanishes above some critical
temperature. It would be very interesting to look for an
experimental realization of the model, possibly in a system
containing a certain conjugated central part, which exhibits the
attractive correlations of carriers with large degeneracy $d>2.$
 Interesting candidate systems are C$_{60}$ molecule
($d=6$) where the electron-phonon interaction is strong
\cite{par}, short nanotubes or other fullerenes ($d \gg 1$), and
mixed-valence molecular complexes. Switching should be fast,
$10^{-13}$ s or faster.

This work has been  supported by DARPA, and by the Leverhulme
Trust (UK).

\end{document}